\documentclass[pre,showpacs,twocolumn]{revtex4}
\usepackage{color}
\usepackage[usenames,dvipsnames]{xcolor}

\usepackage{graphicx}
\usepackage{color}
\usepackage{epsfig}
\usepackage{dcolumn}
\usepackage{bm}
\usepackage{mathrsfs}
\usepackage{multirow}
\usepackage[all]{xy}
\usepackage{pbox}
\usepackage{amssymb}

\def\(({\left(}
\def\)){\right)}
\def\[[{\left[}
\def\]]{\right]}

\newcommand{\beq}{\begin{equation}}
\newcommand{\eeq}{\end{equation}}
\newcommand{\be}{\begin{equation}}
\newcommand{\ee}{\end{equation}}

\newcommand{\la}{\langle}
\newcommand{\ra}{\rangle}

\begin{document}  

\title{Dynamical phase transition for current statistics in a simple driven diffusive system} 

\author{Carlos P. Espigares}
\email[]{cpespigares@onsager.ugr.es}
\affiliation{Departamento de Electromagnetismo y F\'{\i}sica 
de la Materia, and Institute Carlos I for Theoretical and Computational Physics, Universidad de Granada, Granada 18071, Spain}

\author{Pedro L. Garrido}
\email[]{garrido@onsager.ugr.es}
\affiliation{Departamento de Electromagnetismo y F\'{\i}sica 
de la Materia, and Institute Carlos I for Theoretical and Computational Physics, Universidad de Granada, Granada 18071, Spain}

\author{Pablo I. Hurtado}
\email[]{phurtado@onsager.ugr.es}
\affiliation{Departamento de Electromagnetismo y F\'{\i}sica 
de la Materia, and Institute Carlos I for Theoretical and Computational Physics, Universidad de Granada, Granada 18071, Spain}

\date{\today}

\begin{abstract}
We consider fluctuations of the time-averaged current in the one-dimensional weakly-asymmetric exclusion process on a ring. The optimal density profile which sustains a given fluctuation exhibits an instability for low enough currents, where it becomes time-dependent. This instability corresponds to a dynamical phase transition in the system fluctuation behavior: while typical current fluctuations result from the sum of weakly-correlated local events and are still associated with the flat, steady-state density profile, for currents below a critical threshold the system self-organizes into a macroscopic jammed state in the form of a coherent traveling wave, that hinders transport of particles and thus facilitates a time-averaged current fluctuation well below the average current. We analyze in detail this phenomenon using advanced Monte Carlo simulations, and work out macroscopic fluctuation theory predictions, finding very good agreement in all cases. In particular, we study not only the current large deviation function, but also the critical current threshold, the associated optimal density profiles and the traveling wave velocity, analyzing in depth finite-size effects and hence providing a detailed characterization of the dynamical transition.
\end{abstract}


\maketitle

\section{Introduction}
\label{s1}

Recent years are witnessing a quiet revolution in nonequilibrium statistical physics. At the core of this revolution is the realization of the essential role played by macroscopic fluctuations to understand the nonequilibrium behavior of a system of interest \cite{GC,ECM,Ku,LS,Jarzynski,Crooks,HS,Clausius,PT,Bertini}. This activity has led to a number of groundbreaking results valid arbitrarily far from equilibrium (and therefore not restricted to the confining world of linear response), which are offering a glimpse of the long-sought general theory of nonequilibrium phenomena. A main example is the Gallavotti-Cohen fluctuation theorem \cite{GC,ECM,Ku,LS}, which expresses the subtle but enduring consequences of microscopic time reversibility at the macroscopic level. The list continues however, with further breakthroughs ranging from the Jarzynski equality \cite{Jarzynski} or the Crooks fluctuation theorem \cite{Crooks} to the Hatano-Sasa relation \cite{HS} or the recent extension of Clausius inequality to nonequilibrium steady states \cite{Clausius}, to mention just a few \cite{PT}. In addition, a general theoretical framework, the macroscopic fluctuation theory of Bertini and coworkers \cite{Bertini}, has been developed to understand the fluctuating behavior of diffusive systems far from equilibrium (with recent generalizations to driven dissipative media \cite{PLyH,BL}).

A general observation underlying many of these results is that macroscopic fluctuations are often associated with a nontrivial and well-defined path in phase space, a path that the system traverses in order to facilitate such fluctuation. The properties of these optimal paths are revealing a whole new phenomenology at the fluctuating level with important implications out of equilibrium \cite{Pablo1,IFR,Luchinsky}. For instance, the optimal path leading to a macroscopic fluctuation in a nonequilibrium steady state has been recently shown to be the time-reversal of the relaxation path from this fluctuation according to some \emph{adjoint} hydrodynamic laws (which are not necessarily equal to the forward-in-time hydrodynamics) \cite{Bertini}. This general result valid arbitrarily far from equilibrium reduces to the well-known Onsager's regression hypothesis when small deviations from equilibrium are considered. Moreover, the study of the symmetry properties of the optimal paths for current fluctuations has led to another remarkable insight, the isometric fluctuation relation \cite{IFR}, which in turn implies a set of hierarchies of equations for the current cumulants and the nonlinear response coefficients, going far beyond Onsager reciprocity relations and Green-Kubo formulas. 

Another recent and striking discovery concerns the existence of \emph{coherent structures} associated to large, rare fluctuations \cite{BD,PabloSSB}, which in turn imply that these events are far more probable than previously anticipated. Such coherent, self-organized patterns emerge via a dynamical phase transition at the fluctuating level, which is accompanied by spontaneous symmetry breaking \cite{BD,PabloSSB}. The aim of this paper is to investigate in detail this phenomenon in a simple diffusive system in one dimension, namely the weakly-asymmetric simple exclusion process (WASEP) \cite{wasep}, where we study fluctuations of the time-averaged current. 

\begin{figure}
\centerline{\includegraphics[width=9cm,clip]{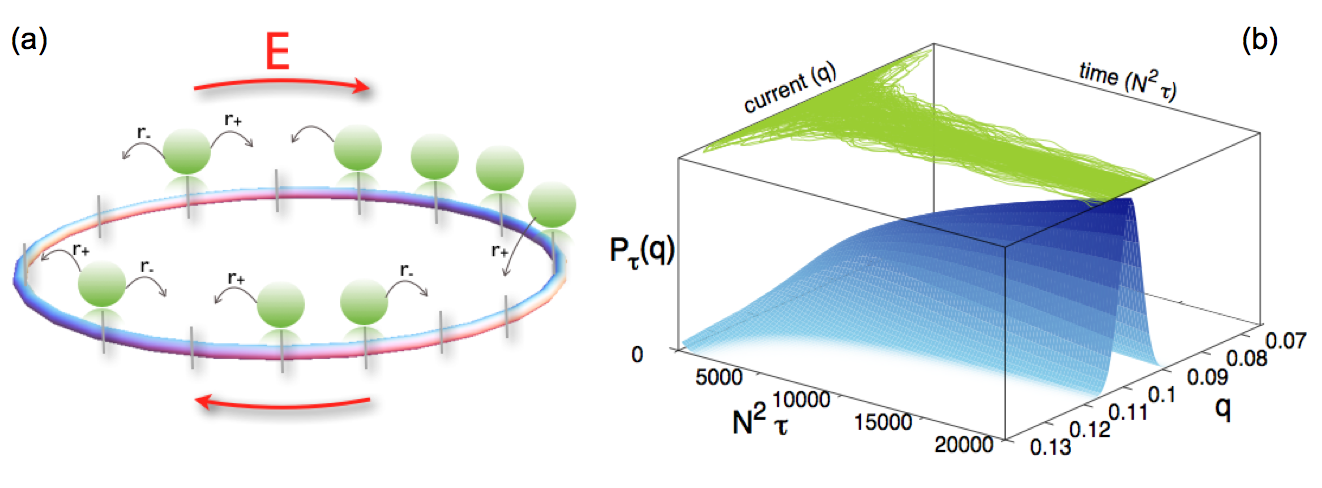}}
\caption{(Color online) (a) Sketch of the WASEP. Particles in a periodic 1d lattice jump stochastically to a right (left) empty nearest neighbor at a rate $r_+$ ($r_-$), so particles feel an external driving field $E=\frac{N}{2}\ln(\frac{r_+}{r_-})$. (b) Convergence of the time-averaged current to its ensemble value $\la q\ra$ for many different realizations (top line cloud), and sketch of the probability concentration as time increases, associated with the large deviation principle, eq. (\ref{ldp}).}
\label{sketch}
\end{figure}

The model is defined on a one-dimensional (1d) lattice of size $N$ with periodic boundary conditions (pbc), where $M\le N$ particles live, see Fig. \ref{sketch}.a, so the total density is $\rho_0=M/N$. Each lattice site $i\in[1,N]$ may contain at most one particle, so the state of the system is defined by a set of occupation numbers, $\mathbf{n}\equiv \{n_i=0,1, \, i\in[1,N]\}$, and $M=\sum_{i=1}^N n_i$. Dynamics is stochastics and proceeds via sequential particle jumps to nearest neighbor sites, provided these are empty, at a rate $r_{\pm}\equiv \frac{1}{2}\exp(\pm E/N)$ for jumps along the $\pm\hat{x}$-direction \cite{note1}. Here $E$ plays the role of a weak external field which drives the system to a nonequilibrium steady state characterized by a homogeneous density profile $\la \rho(x)\ra=\rho_0$ and a nonzero net average current $\la q\ra = \rho_0(1-\rho_0) E$. We employ continuous-time Markov dynamics, so the time to exit a configuration $\mathbf{n}$ is a random variable drawn from a Poisson distribution with exit rate $R(\mathbf{n})= M_+r_+ + M_-r_-$, where $M_{\pm}$ is the number of particles with empty nearest neighbor in the $\pm\hat{x}$-direction.

We are interested in the statistics of the total particle current $q$ flowing through the system, averaged over a long diffusive time $\tau$ and across space. In particular, we define the empirical time-averaged current as $q=t^{-1}(Q_{t}^+-Q_{t}^-)$, where $Q_{t}^{\pm}$ is the total number of particle jumps in the $\pm\hat{x}$-direction in a given microscopic time interval $t=\tau N^2$.  For $\tau \to \infty$ this time-averaged estimate converges toward the ensemble average $\la q\ra$. However, for long but finite times $\tau$ we observe fluctuations $q\ne \la q\ra$, and their probability $\text{P}_{\tau}(q)$ obeys a large deviation principle in this limit \cite{LD}
\be
\text{P}_{\tau}(q)\sim \text{e}^{+\tau N G(q)} \, ,
\label{ldp}
\ee
where $G(q) \le 0$ is the current large deviation function (LDF), such that $G(\la q\ra)=0$.
This means that the probability of observing a fixed current fluctuation $q\neq \la q \ra$ decays exponentially as both $\tau$ and $N$ increase, at a rate given by $G(q)$, see Fig. \ref{sketch}.b. In other words, $G(q)$ measures the rate at which $P_{\tau}(q)$ concentrates around $\la q\ra$. The above large-deviation principle describes the scaling of $\text{P}_{\tau}(q)$ for both typical and rare current fluctuations. In particular, a suitable expansion of $G(q)$ for small fluctuations yields the usual gaussian form for $\text{P}_{\tau}(q)$ associated with the central limit theorem. The current LDF plays an important role in nonequilibrium statistical physics as it contains essential information on the transport properties of the system at hand. Moreover, in general LDFs play in nonequilibrium physics a role akin to the free energy of equilibrium systems. Therefore, even though we do not know how to connect in general microscopic dynamics to macroscopic properties in nonequilibrium systems (in a way equivalent to the equilibrium ensemble formalism), we can still measure LDFs of macroscopic observables out of equilibrium, which provide an alternative path to a detailed macroscopic description of nonequilibrium phenomena.

\section{Macroscopic fluctuation theory and dynamical phase transition}
\label{s2}

\begin{figure}
\centerline{\includegraphics[width=9cm,clip]{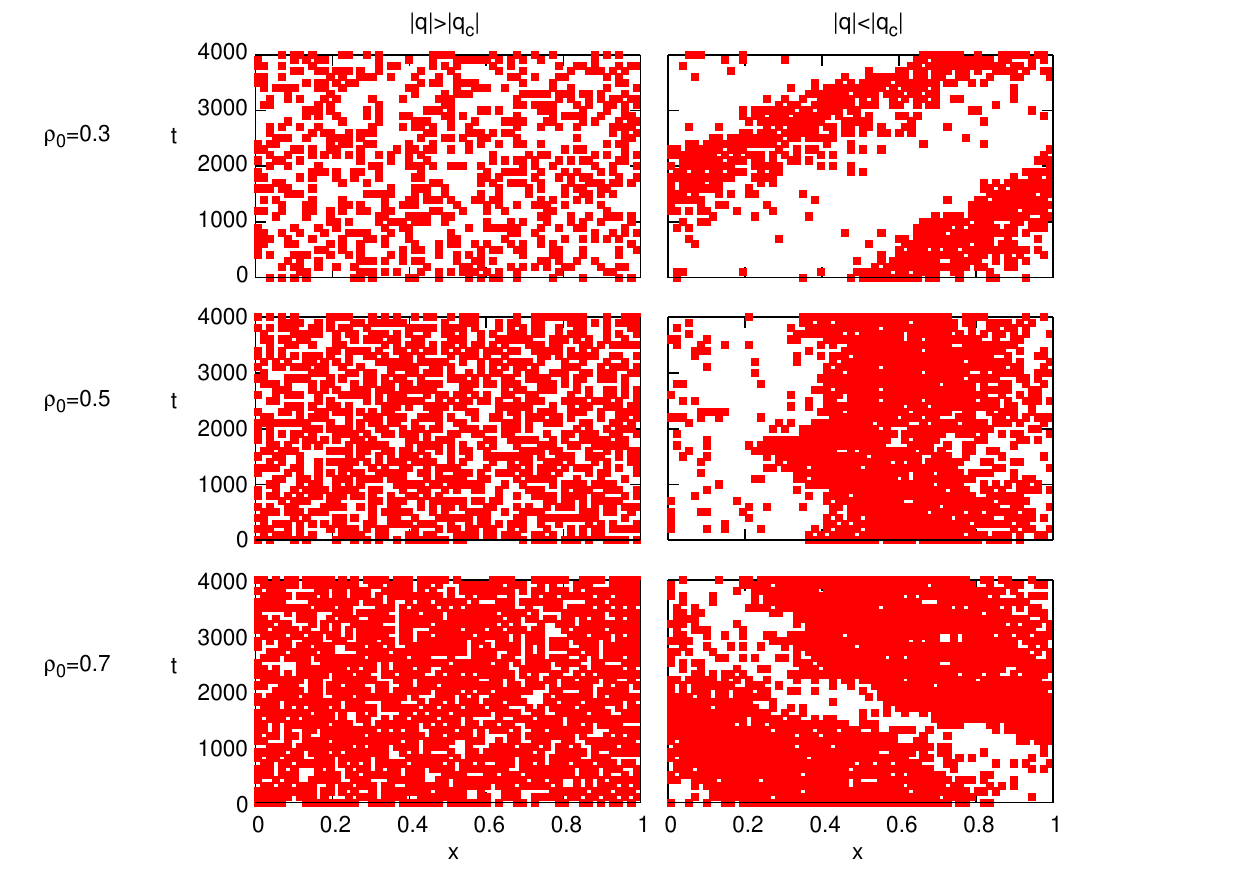}}
\caption{(Color online) Typical evolution of microscopic configurations for current fluctuations above and below the critical current for three different densities in the WASEP. 
Left panels correspond to currents above the critical one where the system remains homogeneous. Right panels correspond to subcritical current fluctuations where a traveling wave emerges. The velocity of the traveling wave of the top right panel ($\rho_0=0.3$) is positive. The {\it{traveling}} wave of the central right panel does not move on average, corresponding to $\rho_0=1/2$, and the wave at the bottom left panel ($\rho_0=0.7$) moves with negative velocity.
}
\label{evolution}
\end{figure}

Computing LDFs from scratch, starting from microscopic dynamics, is a humongous task which has been achieved only in a handful of oversimplified models (most of them stochastic lattice gases and related models) \cite{Bertini,wasep}. However, in a recent series of works, Bertini and collaborators \cite{Bertini} have developed a phenomenological theory, the macroscopic fluctuation theory (MFT), which describes in detail dynamic fluctuations in driven diffusive systems starting from the hydrodynamic evolution equation for the local density $\rho(x,t)$ for the system of interest and the sole knowledge of two transport coefficients, the diffusivity $D(\rho)$ and the mobility $\sigma(\rho)$, which can be measured experimentally. From this knowledge, MFT offers explicit predictions for the current LDF (see the Appendix)
\be
G(q) = -\frac{1}{\tau} \min_{\{\rho,j\}_0^{\tau}}\int_0^{\tau} dt \int_0^1 dx \frac{[j+D(\rho)\partial_x\rho-\sigma(\rho) E]^2}{2\sigma(\rho)} \, ,
\label{ldfq}
\ee
where a long-time limit is implicit and the minimum is taken over all histories of the density and current fields, $\rho(x,t)$ and $j(x,t)$ respectively, coupled via the continuity equation $\partial_t \rho + \partial_x j = 0$ at every point of space and time, and subject to the constraint $q=\tau^{-1}\int_0^{\tau} dt \int_0^1 dx j(x,t)$ for the space\&time-averaged current and the appropriate boundary conditions (periodic in this case), see Appendix \ref{append} for details. Note that, for the WASEP, $D(\rho)=\frac{1}{2}$ and $\sigma(\rho)=\rho(1-\rho)$ \cite{wasep}. The optimal density and current fields solution of the above variational problem, denoted here as $\rho_q(x,t)$ and $j_q(x,t)$,  can be interpreted as the path the system follows in mesoscopic phase space in order to sustain a given current fluctuation $q$. This path may be in general time-dependent, and the resulting general variational problem is remarkably hard. This problem becomes simpler however in different limiting cases. For instance, one expects that small current fluctuations around the average, $q\simeq \la q \ra$, result from the random superposition of weakly-correlated (if any) local fluctuations of the microscopic jump process. In this case one expects the optimal density field to be just the flat, steady-state one, $\rho_q(x,t)=\rho_0$, and hence $j_q(x,t)=q$, resulting in a simple quadratic form for the current LDF
\be
G_{\text{flat}}(q) = - \frac{(q-\sigma(\rho_0) E)^2}{2\sigma(\rho_0)} \, .
\label{gflat}
\ee
Therefore gaussian statistics is obtained for small (i.e. typical) current fluctuations, in agreement with the central limit theorem. The argument above breaks down however for moderate current fluctuations. In fact, Bodineau and Derrida have shown recently \cite{BD} that the flat profile indeed becomes unstable, in the sense that $G(q)$ increases by adding a small time-dependent periodic perturbation to the otherwise constant profile, whenever $8 \pi^2D^2(\rho_0)\sigma(\rho_0)+(E^2\sigma^2(\rho_0)-q^2)\sigma''(\rho_0)<0$, where $\sigma''$ denotes the second derivative. This condition yields a \emph{critical current}
\be
|q_c|=\sqrt{\frac{8\pi^2D^2(\rho_0)\sigma(\rho_0)}{\sigma''(\rho_0)}+E^2\sigma^2(\rho_0)} \, ,
\label{qc}
\ee
which signals the onset of the instability. This instability can be interpreted as a dynamical phase transition at the fluctuating level, and involves the spontaneous breaking of translation symmetry (see Fig. \ref{evolution}). In fact, for the WASEP the dynamic phase transition corresponds to the emergence of a macroscopic jammed state which hinders transport of particles to facilitate a current fluctuation well below the average. When the instability kicks in, an analysis of the resulting perturbation \cite{BD} suggests that the dominant form of the optimal profile is a traveling wave, $\rho_q(x,t)=\omega_q(x-vt)$, moving at constant velocity $v$ across the system \cite{BD,PabloSSB}. Provided that the traveling-wave form remains as the optimal solution for currents well-below the critical threshold, the current LDF can now be written as
\begin{eqnarray}
G(q)=-\min_{\omega_q(x),v}\int_0^1\frac{dx}{2\sigma[\omega_q(x)]}[q-v\rho_0+v\omega_q(x)
\nonumber \\
+D[\omega_q(x)]\omega'_q(x)-\sigma[\omega_q(x)]E]^2 \, ,
\label{ldfwave}
\end{eqnarray}
where the minimum is now taken over the traveling wave profile $\omega_q(x)$ and its velocity $v$. Notice that, for the instability to exist, the strength of the driving field, $|E|$, must be large enough to guarantee a positive discriminant in eq. (\ref{qc}), namely
\be
|E|\geq |E_c| \equiv \text{Re}\left[\sqrt{-\frac{8\pi^2D(\rho_0)^2}{\sigma(\rho_0)\sigma''(\rho_0)}}\right] \, .
\label{Ec}
\ee
Therefore, since the mobility $\sigma(\rho)$ is positive definite, a non-zero threshold field only exists for models such that $\sigma''(\rho)<0$, which is the case of the WASEP here studied, where $\sigma(\rho)=\rho(1-\rho)$. Other transport models, as for instance the Kipnis-Marchioro-Presutti (KMP) model of heat conduction \cite{kmp,PabloSSB}, have $\sigma''(\rho)>0$ and hence $|E_c|=0$, thus exhibiting the aforementioned instability even in the absence of external fields \cite{PabloSSB}.

It is worth noting that MFT inherits the microscopic symmetries of the system of interest. In our particular case, the WASEP shows a clear particle-hole symmetry, and this is reflected in the current LDF, above and below the instability. In particular, the optimal wave profile $\omega_q(x)$, associated with a current fluctuation $|q|<|q_c|$ for a density $\rho_0$, is complementary to the optimal wave profile for the same value of $q$ and density $1-\rho_0$, i.e.
\beq
\omega_{q}(x;\rho_0)=1-\omega_{q}(x;1-\rho_0).
\label{profrel}
\eeq
In addition, the optimal wave for density $1-\rho_0$ travels with the same speed but opposite direction to that of the corresponding wave for density $\rho_0$, i.e. $v_q(\rho_0)=-v_q(1-\rho_0)$. In the particular case of $\rho_0=1/2$ these relations imply that, for any current fluctuation, the optimal density profile and its complementary are equivalent, and the velocity of the optimal traveling wave is zero for all fluctuations. Therefore, for $\rho_0=1/2$ the typical macroscopic configurations in the time-dependent regime have a well defined wave structure which however does not move on average. This can be observed in Fig. \ref{evolution}, where typical system space-time trajectories for current fluctuations above and below the critical current are displayed for $\rho_0=0.3,\, 0.5,$ and $0.7$.  Notice that for $|q|<|q_c|$ there is a nontrivial structure which travels with opposite velocities for $\rho_0=0.3$ and $0.7$, and which does not move when $\rho_0=0.5$. Furthermore, using the above symmetry relations for the WASEP current LDF we find that
\beq
G(q;\rho_0)=G(q;1-\rho_0) \, .
\label{LDFrel}
\eeq
Hence, given a external field, it is enough to compute the current LDF for $\rho_0\in [0,1/2]$. 

Another interesting symmetry, though far less obvious, is related to the time-reversibility of microscopic dynamics. This relation, known as the Gallavotti-Cohen fluctuation theorem \cite{GC,ECM,Ku,LS}, implies a remarkably simple connection between the probability of a given current fluctuation $q$ and the reverse event, $-q$, which can be stated for the current LDF in the following way
\be
G(q)-G(-q)=2E q \, .
\label{GC}
\ee
This in turn implies that the odd part of the typically nontrivial function $G(q)$ is \emph{linear} in the current, with an universal coefficient $2E$. This symmetry can be also stated for the Legendre transform of the current LDF, see eq. (\ref{mu}) below
\be
\mu(\lambda)=\mu(-\lambda-2E) \, .
\label{GCmu}
\ee
As we will see below, this fluctuation relation is fully confirmed in our simulations, both below and above the instability. Moreover, the Gallavotti-Cohen relation can be used to bound the validity of the simulation method used to explore large deviations \cite{Pablo1,Pablo2}, see below. 

\begin{figure}
\centerline{\includegraphics[width=8.cm,clip]{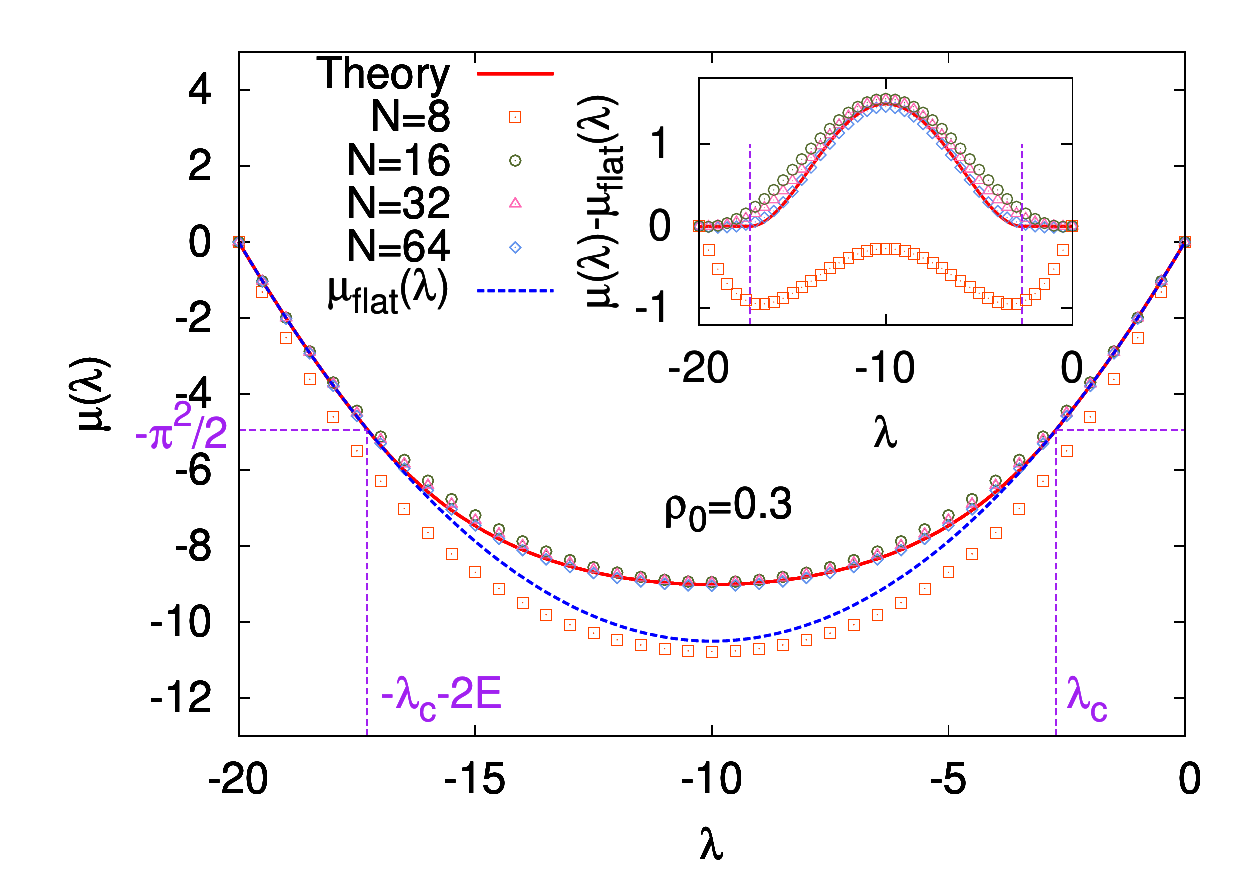}}
\centerline{\includegraphics[width=8.cm,clip]{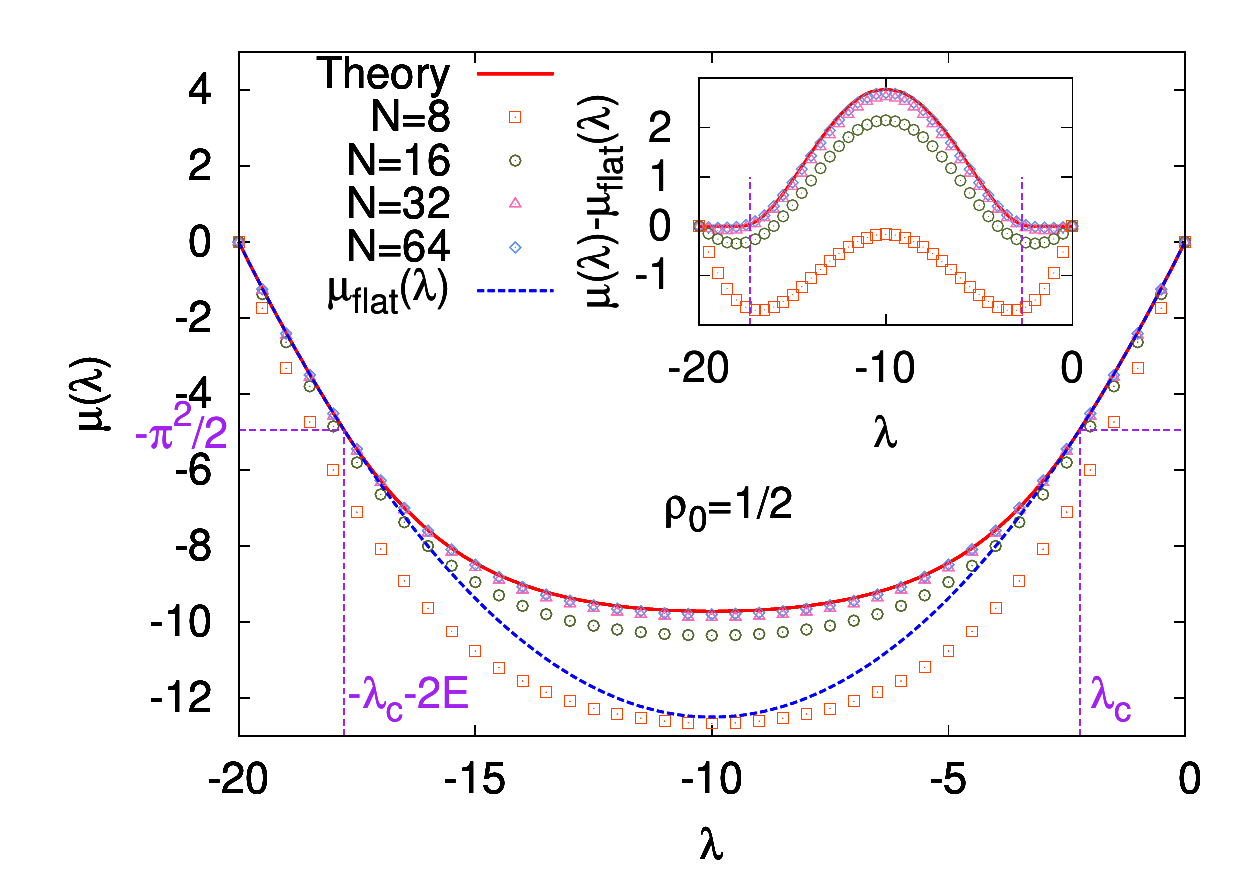}}
\caption{(Color online) Legendre transform of the current LDF, $\mu(\lambda)$. Top: Measured $\mu(\lambda)$ for $\rho_0=0.3$ and increasing $N$, together with the MFT result (solid red line) and the quadratic approximation (dashed blue line). Bottom: Equivalent data for $\rho_0=1/2$. Insets: $\mu(\lambda)-\mu_{\text{flat}}(\lambda)$ for the same $N$ and $\rho_0=0.3$ (top) and $0.5$ (bottom). In all cases, data converge to the MFT prediction as $N$ increases.}
\label{mu_lamb}
\end{figure}

\section{Numerical results}
\label{s3}

Our aim in this paper is to characterize in detail the dynamical phase transition in current statistics for the WASEP using numerical simulations, and compare with the predictions derived within MFT. These predictions are explicitly worked out in the Appendix. The critical current and the threshold field for the WASEP are
\begin{eqnarray}
|q_c| & = & \rho_0(1-\rho_0)\sqrt{E^2 - E_c^2} \, \label{qc2} \\
|E_c| & = & \frac{\pi}{\sqrt{\rho_0(1-\rho_0)}} \, .
\label{Ec}
\end{eqnarray}
Equivalently, taking into account that $\la q\ra = \rho_0(1-\rho_0) E$, we have that 
\be
|q_c| = |\la q\ra | \sqrt{1-\left(\frac{E_c}{E}\right)^2} \, ,
\label{qcWASEP}
\ee
so $|q_c| < |\la q\ra |$ in WASEP for field strengths above the threshold, while no phase transition happens for $|E|<|E_c|$.

In order to investigate the instability described in Section \ref{s2} using numerical simulations, we need to explore the statistics of both typical and rare current fluctuations. While the former pose no problem and can be studied in standard simulations, to sample the atypical trajectories associated with rare current fluctuations we must resort to advanced Monte Carlo methods that allow to measure directly LDFs in many particle systems \cite{sim,sim2,sim3}. 
This technique implies a modification of the stochastic microscopic dynamics, in such a way that the rare events responsible of a large current fluctuation are no longer rare with the modified dynamics. The numerical method also requires the parallel simulation of multiple \emph{clones} or copies of the system \cite{sim,sim2,sim3}, which may be replicated or pruned depending on its importance for the particular fluctuation we want to measure. In this work we used in particular $N_c=2\times 10^4$ clones for $\rho_0=0.3$ and $N_c=5\times 10^4$ for $\rho_0=1/2$, and we checked that results do not depend on the total number of clones for large enough $N_c$ \cite{Pablo2}. The method yields a Monte Carlo estimate of the Legendre transform of the current LDF
\be
\mu(\lambda)=\max_q[G(q) + \lambda q] \, , 
\label{mu}
\ee
with $\lambda$ a parameter conjugated to the current, such that $G'(q)+\lambda=0$. In this way, the function $\mu(\lambda)$ can be seen as the conjugate \emph{potential} to $G(q)$, a relation equivalent to the free energy being the Legendre transform of the internal energy in thermodynamics, with the temperature as conjugate parameter to the entropy. Legendre-transforming the quadratic current LDF in eq. (\ref{gflat}) once particularized for WASEP, obtained for the time-independent (homogeneous) fluctuation regime, we have
\be
\mu_{\text{flat}}(\lambda)=\frac{\rho_0}{2}(1-\rho_0)\lambda (\lambda+2E) \, .
\label{muflat}
\ee
However, for currents in a well-defined interval $|q|<|q_c|$, with $|q_c|$ defined in eqs. (\ref{qc2}) or (\ref{qcWASEP}), a time-dependent regime with an optimal density profile in the form of a traveling wave is expected. This corresponds to values of the conjugate parameter $\lambda$ such that $|\lambda+E| < \Lambda_c$, with
\be
\Lambda_c \equiv \frac{q_c}{\rho_0(1-\rho_0)} = \sqrt{E^2 - \frac{\pi^2}{\rho_0(1-\rho_0)}} = \sqrt{E^2 - E_c^2}\, ,
\label{lambdac}
\ee
and where we have used eqs. (\ref{qc2})-(\ref{qcWASEP}). Therefore we expect traveling wave solutions for $\lambda_c^-<\lambda<\lambda_c^+$, where
\be
\lambda_c^{\pm}\equiv \pm\Lambda_c - E \, .
\label{lambdacpm}
\ee
In this way the time-dependent fluctuation regime kicks in whenever $\mu(\lambda)<\mu(\lambda_c^{\pm})=-\pi^2/2$, see eq. (\ref{muflat}).

\begin{figure}
\centerline{\includegraphics[width=8cm]{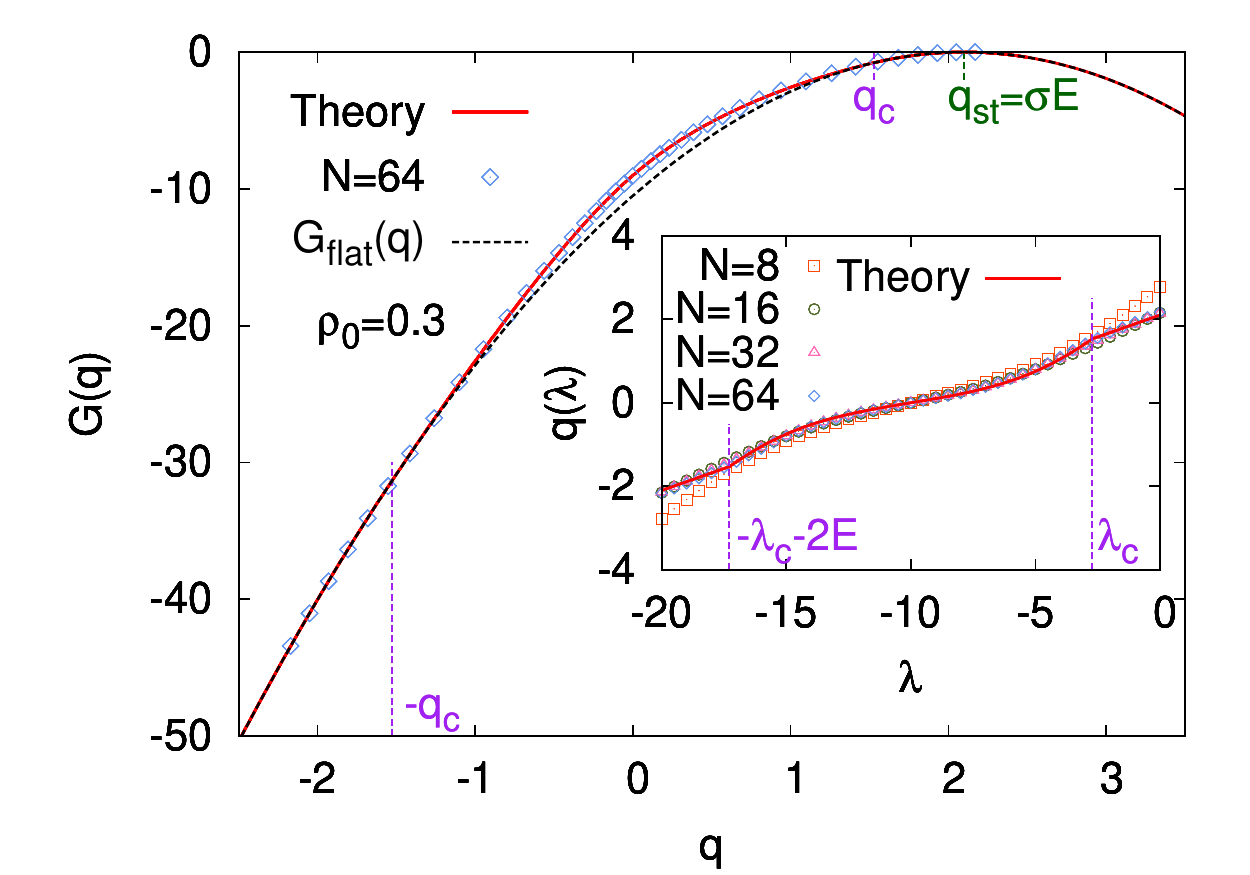}}
\centerline{\includegraphics[width=8cm]{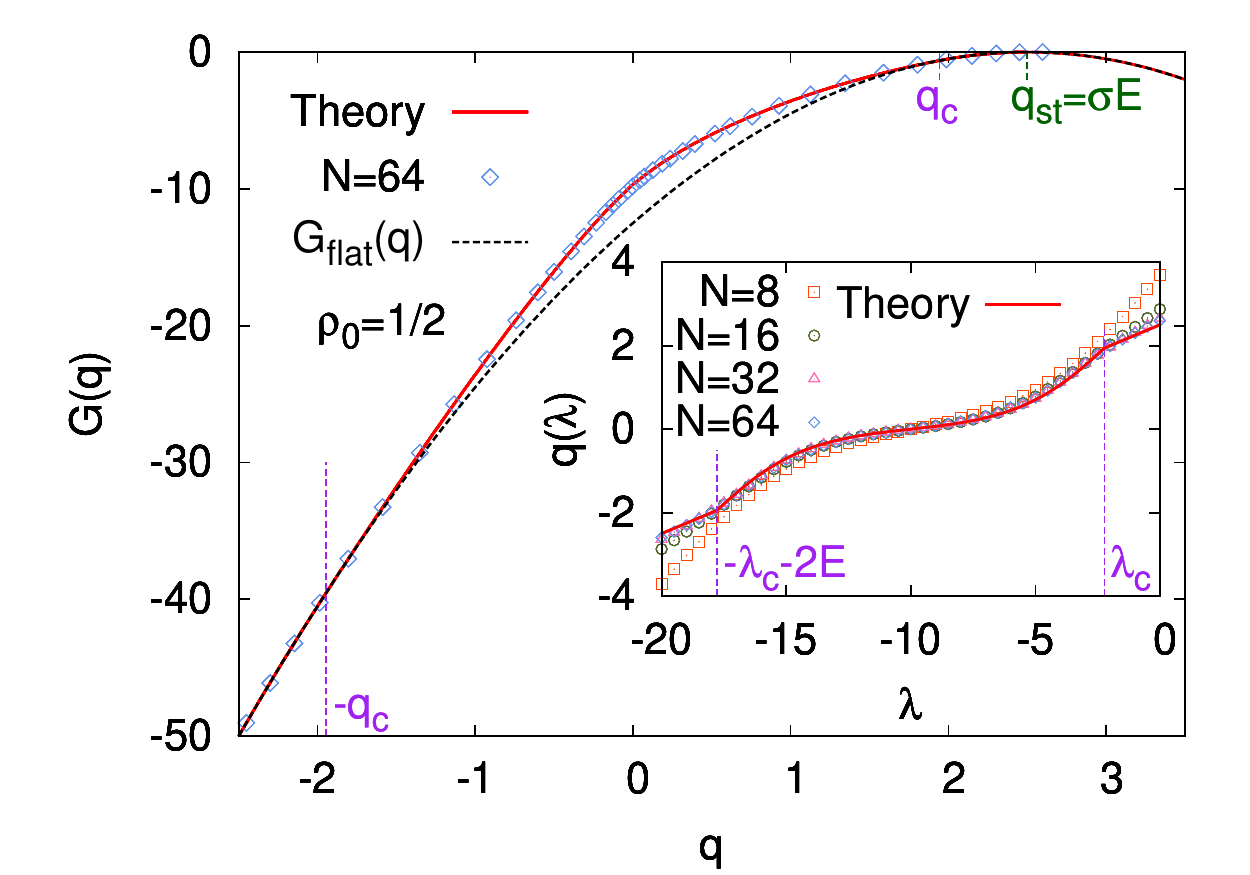}}
\caption{(Color online) Top: Large Deviation Function for $\rho_0=0.3$ Inset: Measured average current $q_{\lambda}$ as a function of $\lambda$ and increasing N, together with the analytical prediction base on the MFT. Bottom: Same results for $\rho_0=1/2$. For both densities the traveling wave solution enhances the probability for fluctuations $|q|<|q_c|$ (solid red line) with respect to the flat profile associated to gaussian statistics (dashed black line).}
\label{G_q}
\end{figure}

We performed simulations of the 1d periodic WASEP for three different average densities, $\rho_0=0.3,\, 0.5$ and $0.7$, for increasing system sizes $N\in[8,64]$ and a fixed external field $E=+10$. This driving field is above the threshold $E_c$ for all three densities $\rho_0$, see eq. (\ref{Ec}), so we expect the instability to appear on the basis of the analysis of Section \ref{s2}. Fig. \ref{mu_lamb} shows simulation results for $\mu(\lambda)$ and increasing values of $N$ for two different values of $\rho_0$, together with the explicit MFT results (see the Appendix). Gaussian current statistics corresponds to a quadratic behavior in $\mu(\lambda)\approx \mu_{\text{flat}}(\lambda)$, see eq. (\ref{muflat}), which is fully confirmed in Fig. \ref{mu_lamb} for $|\lambda+E|>\Lambda_c$ and different values of $N$. This means that small current fluctuations, $q\simeq\la q \ra$, have their origin in the superposition of uncorrelated (or at most weakly-correlated) local events of the stochastic jump process, giving rise to Gaussian statistics as dictated by the central limit theorem and thus confirming their incoherent origin. Interestingly, this observation also applies to the time-reversal partners of these small fluctuations, $-q$, which are far from typical. This is a direct consequence of the Gallavotti-Cohen symmetry \cite{GC,ECM,Ku,LS}, which implies that the statistics associated with a current fluctuation does not depend on the current sign \cite{Pablo2}. On the other hand, for fluctuations below a critical threshold, $|q|<|q_c|$ or equivalently $\lambda_c^-<\lambda<\lambda_c^+$, deviations from this simple quadratic form are apparent, signaling the onset of the dynamical phase transition anticipated in Section \ref{s2}, see also Fig. \ref{evolution}. In fact, as $N$ increases a clear convergence toward the MFT prediction (which is strongly non-quadratic in the regime $\lambda_c^-<\lambda<\lambda_c^+$) is observed, with very good results already for $N=64$. Strong finite size effects associated with the finite population of clones $N_c$ prevent us from reaching larger system sizes \cite{Pablo1,Pablo2,PabloSSB}, but $N=64$ is already \emph{close} enough to the asymptotic hydrodynamic behavior. Still, small corrections to the MFT predictions are observed, see the insets of Fig. \ref{mu_lamb}, which quickly decrease with $N$. On the other hand, the Gallavotti-Cohen fluctuation theorem for currents holds in the whole current range, see eq. (\ref{GCmu}) and Fig. \ref{mu_lamb}, both in the homogeneous and time-dependent current fluctuation regimes. Furthermore, the Gallavotti-Cohen symmetry holds irrespective of $N$, as a result of the microreversibility of the model at hand: while we need a large size limit in order to verify the predictions derived from MFT (which is a macroscopic theory), no finite-size corrections affect the fluctuation theorem, whose validity can be used to ascertain the range of applicability of the cloning algorithm used to sample large deviations \cite{Pablo2}.

\begin{figure*}
\centerline{\includegraphics[width=18cm,clip]{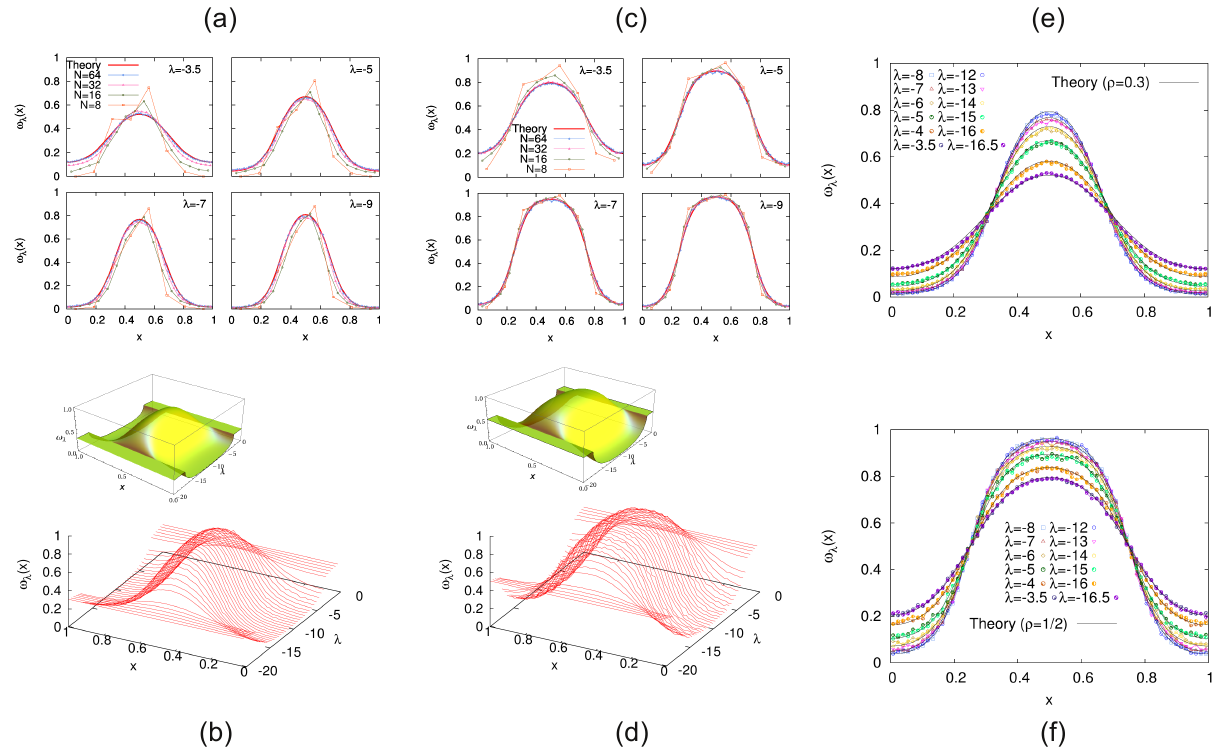}}
\caption{(Color online) Measured traveling wave profiles for different values of $\lambda\in(\lambda_c^-,\lambda_c^+)$, varying $N\in[8,64]$ and two different average densities $\rho_0=0.3, \, 0.5$, together with theoretical predictions from MFT, see the Appendix. (a) $\omega_{\lambda}(x)$ measured for $\rho_0=0.3$, different $\lambda$ and increasing $N$, and MFT predictions. (b) Measured density profiles as a function of $\lambda$ for $N=64$. Optimal profiles are flat up to a critical current (equivalently $\lambda_c^{\pm}$) where a traveling wave emerges. The inset shows the MFT prediction. (c) and (d) are equivalent to (a) and (b), respectively, but for an average density $\rho_0=0.5$. Notice the comparatively \emph{thicker} wave for $\rho_0=0.5$. (e) Collapse of measured profiles associated with different current fluctuations $\omega_{\lambda}(x)$ and their time-reversal partners $\omega_{-\lambda-2E}(x)$ for $N=64$ and $\rho_0=0.3$, together with theoretical predictions. Optimal profiles, both below and above the instability, remain invariant under change of sign of the current. (f) Same results for $\rho_0=0.5$.}
\label{profiles}
\end{figure*}

We also measured the average current $\la q_{\lambda}\ra$ associated with a fixed value of the conjugate parameter $\lambda$. The insets of Fig. \ref{G_q} show our results and the predictions based on MFT. Again the agreement is excellent (improving as $N$ increases), both above and below the dynamical phase transition, even though there is a clear change of behavior across the transition points $\lambda_c^{\pm}$. In particular, in the Gaussian fluctuation regime the current is linear in $\lambda$, namely $\la q_{\lambda}\ra = \rho_0(1-\rho_0)(\lambda+E)$, while the relation becomes strongly non-linear in the time-dependent region, $|\lambda+E| < \Lambda_c$. We may use the measured $\la q_{\lambda}\ra$ to give a direct Monte Carlo estimate of the current LDF $G(q)$. In fact, $\la q_{\lambda}\ra$ is the current conjugated to a given $\lambda$ and hence we may write $G(q)=\mu(\lambda)+\lambda \la q_{\lambda}\ra$, where we combine the measured $\mu(\lambda)$ in Fig. \ref{mu_lamb} and the measured $\la q_{\lambda}\ra$ in the insets of Fig. \ref{G_q}. The result for $G(q)$ and different values of $\rho_0$ is plotted in Fig. \ref{G_q}, where we again find a good agreement between theory and Monte Carlo simulation. Notice in particular the deviation from quadratic behavior observed for currents $|q|<|q_c|$ resulting from the formation of macroscopic jammed states (see below). 

The dynamical phase transition is most evident at the configurational level, as observed in Fig. \ref{evolution}, so we measured the average density profile associated with a given current fluctuation \cite{Pablo1} for different values of the total density $\rho_0$; see Fig. \ref{profiles}. Because of the system periodicity, and in order not to blur away the possible structure present in microscopic configurations, we performed profile averages around the instantaneous center of mass. For that, we consider the system as a 1d ring embedded in two-dimensional space, see Fig. \ref{sketch}.a,  and compute the angular position of the center of mass, shifting it to the origin before averaging. In particular, we assign an angular position $\theta_i=2\pi i/N$ to each site $i\in[1,N]$ in the lattice. The angular position of the center of mass for a given microscopic configuration $\mathbf{n}=\{n_i,i=1,\ldots,N\}$, with $n_i=0,1$ the on-site occupation numbers, is thus defined as
\be
\theta_{\text{CM}} \equiv  \tan^{-1}\left(\frac{Y_{\text{CM}}}{X_{\text{CM}}}\right)
\label{angCM}
\ee
where
\be
X_{\text{CM}}= \frac{1}{M} \sum_{i=1}^N n_i \cos \theta_i \quad ; \quad Y_{\text{CM}}= \frac{1}{M} \sum_{i=1}^N n_i \sin \theta_i \, ,
\label{xyCM}
\ee
and recall that $M=\sum_{i=1}^N n_i$ is the total number of particles. Notice that this center-of-mass averaging procedure yields a spurious weak structure in the Gaussian (homogeneous) fluctuation region, equivalent to averaging random  particle profiles around their (random) center of mass. Such a spurious profile is of course independent of the current $q$ and can be easily subtracted. On the other hand, once the instability is triggered average profiles exhibit a much more pronounced structure resulting from the appearance of a traveling wave; see right column in Fig. \ref{evolution}. Figures \ref{profiles}.a,c show the measured profiles $\omega_{\lambda}(x)$ for different $\lambda\in(\lambda_c^-,\lambda_c^+)$, varying $N\in[8,64]$ and two different average densities $\rho_0=0.3$ (a) and $0.5$ (c), together with theoretical predictions from MFT as calculated in the Appendix. Again, fast convergence toward the MFT result is observed, with excellent agreement for $N=64$ in all cases. Moreover, Figs. \ref{profiles}.b,d show a three-dimensional plot of the measured profiles for $N=64$ and different $\lambda$, again for $\rho_0=0.3$ (b) and $0.5$ (d), which closely resembles the MFT scenario plotted as insets to these figures. In general, the traveling wave profile grows from the flat form as $\lambda$ crosses the critical values $\lambda_c^{\pm}$ penetrating into the critical region, thus favoring a macroscopic jammed state that hinders transport of particles and thus facilitates a time-averaged current fluctuation well below the average current. This macroscopic jammed state reaches its maximum expression for $\lambda=-E$, or equivalently $q=0$ --see the insets of Fig. \ref{G_q}, so the system is maximally jammed for zero current irrespective of the driving field $E$. 

\begin{figure}
\centerline{\includegraphics[width=8.cm,clip]{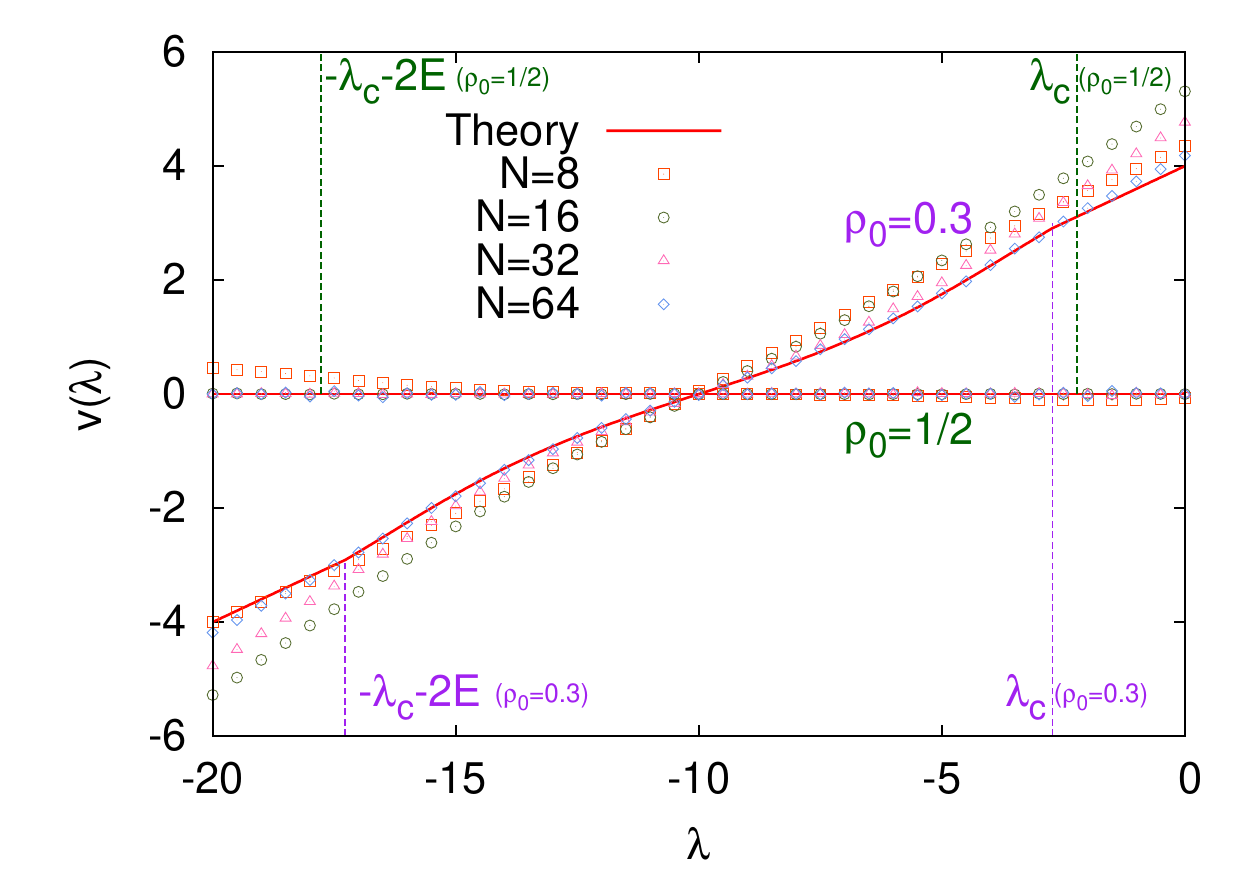}}
\caption{(Color online) Measured velocity as a function of $\lambda$ for $\rho_0=0.3$, $\rho_0=1/2$ and increasing N, together with the 
MFT result.}
\label{v_lamb}
\end{figure}

An interesting corollary of the Gallavotti-Cohen fluctuation symmetry \cite{GC,ECM,Ku,LS} is that the optimal density profile associated with a given current fluctuation remains invariant under changes of the current sign, i.e. $\omega_q(x)=\omega_{-q}(x)$, independently of the driving external field \cite{Pablo1,Pablo2}. We confirm this property in Fig. \ref{profiles}.e,f, where we plot for $N=64$ and different $\rho_0$ the optimal density profiles for different pairs of  fluctuations coupled by time-reversal, i.e. for pairs of values $(\lambda,-\lambda-2E)$, finding an excellent collapse as predicted by theory. Moreover, the collapsing pairs of profiles agree to a high degree of accuracy with the theoretical curves.

We also measured the average velocity associated with a given current fluctuation by fitting the motion of the center of mass during small time intervals $\Delta t$ to a ballistic law, $r_{\text{cm}}(t+\Delta t) - r_{\text{cm}}(t) = vt$, see, e.g., right column in Fig. \ref{evolution}, and making statistics for the measured velocity. Figure \ref{v_lamb} shows the mean velocity for $\Delta t=200$ Monte Carlo steps as a function of $\lambda$ for increasing values of $N$ and different values of $\rho_0$, and the agreement with MFT is again very good already for $N=64$ (we checked that other values of $\Delta t$ yield equally good results). As for the current, see the insets of Fig. \ref{G_q}, there is a clear change of tendency across the instability points $\lambda_c^{\pm}$, with the velocity being a linear function of $\lambda$ in the Gaussian regime but turning strongly nonlinear in the traveling-wave region. Interestingly, the measured wave average velocity for $\rho_0=0.5$ is compatible zero for all current fluctuations, hence confirming the MFT prediction based on the particle-hole symmetry of WASEP.

\section{Conclusions}
\label{s4}

In this paper we have considered the statistics of the time-averaged current in a simple driven diffusive system, the 1d weakly asymmetric simple exclusion process on a ring. This system exhibits a dynamical phase transition at the fluctuation level for large enough driving external fields, and we have characterized such instability in detail using advanced numerical simulations. We show in particular that typical (i.e. small) current fluctuations result from the sum of weakly-correlated local events which take place in an otherwise homogeneous density background on average, and thus give rise to Gaussian current statistics in agreement with the central limit theorem. However, for large enough current fluctuations well below the average current (namely in a well-defined range of current fluctuations around zero current), the system breaks this homogeneity and self-organizes its density profile into a macroscopic jammed state in the form of a coherent traveling wave moving at constant velocity. This wave structure hinders transport of particles and thus facilitates a time-averaged current fluctuation well below the average current. It is worth emphasizing that the emergence of a traveling wave breaks spontaneously a symmetry (translation invariance) at the fluctuating level. This phenomenon is fully captured by macroscopic fluctuation theory, whose predictions are explicitly worked out and confirmed to a high degree of accuracy by simulation results. Our study offers insights not only on the current large deviation function which characterizes current statistics, but also on the optimal density profiles associated with the different fluctuations and their velocity, as well as the effect of finite-size corrections on the observables of interest, providing a detailed characterization of the dynamical transition not available before.

A similar dynamical phase transition has been recently observed and characterized in another model of transport, the Kipnis-Marchioro-Pressuti (KMP) model of heat conduction \cite{PabloSSB}. In that case the phenomenon is even more striking, as it happens in an isolated system in the absence of any external field, spontaneously breaking a symmetry in 1D and illustrating the idea that critical phenomena not allowed in equilibrium steady states may however arise in their fluctuating behavior or under nonequilibrium conditions. Although both instabilities are described equally well by MFT, the physical interpretation of the dynamical transition is quite different. In particular, in the KMP model the instability happens because the system optimizes the transport of a \emph{large} current by gathering energy in a localized packet (the wave) which then travels coherently, breaking spontaneously translation symmetry in the process. On the other hand, the WASEP instability happens in order to hinder the transport of particles via the formation of a macroscopic jammed states (the wave), thus facilitating a current fluctuation well \emph{below} the average. Interestingly, both phenomena are sides of essentially the same instability. It is also worth noticing that similar instabilities have been described in quantum systems \cite{quantum}. 

Our results show unambiguously that the dynamical phase transition observed in the WASEP is continuous as conjectured in \cite{BD}, excluding the possibility of a first-order scenario, in concordance with previous observations for the KMP model \cite{PabloSSB}. This suggests that a traveling wave is in fact the most favorable time-dependent profile once the instability is triggered. This observation may greatly simplify general time-dependent calculations, but the question remains of whether this is the whole story or if other, more complex solutions may play a dominant role for even larger fluctuations. An interesting, related question concerns the properties of time-dependent solutions for systems with open boundaries, where traveling-wave patterns are not appropriate. The time-independent profiles in these cases, from which a suitable perturbation analysis would hint at the form of the time-dependent solution, are far more complex than the trivial homogeneous profiles that appear for periodic systems, difficulting progress along this line. In fact, a recent study \cite{Gorissen} has found no evidence of dynamical phase transition in WASEP with open boundaries. In any case, it seems clear that extremely rare events call in general for coherent, self-organized patterns in order to be sustained \cite{rare}.

Another interesting direction to explore in a near future is the appearance of this phenomenon in higher-dimensional systems. In this case the solution of the associated MFT is far more complicated, with no guarantee of an unique solution and several competing patterns already known \cite{Carlos}. The role of numerical simulation will hence prove essential to explore rare current fluctuations in high-dimensional systems and to understand the appearance of dynamical phase transitions at the fluctuation level \cite{Carlos}. Furthermore, the simplicity and elegance of this phenomenon suggests that it might be a rather general property of any fluctuating field theory, with possible expressions in quantum field theory, hydrodynamics, etc. 

\begin{acknowledgements}
Financial support from Spanish MICINN project FIS2009-08451, University of Granada, and Junta de Andaluc\'{\i}a projects P07-FQM02725 and P09-FQM4682 is acknowledged.
\end{acknowledgements}

\appendix*

\section{Macroscopic fluctuation theory for current statistics and dynamical phase transition}
\label{append}

Macroscopic fluctuation theory (MFT) \cite{Bertini} describes in detail dynamic fluctuations in driven diffusive systems, starting from the hydrodynamic evolution equation for the system of interest and the sole knowledge of two transport coefficients, which can be measured experimentally. From this knowledge, MFT offers explicit predictions for the current LDF and the associated path in phase space responsible of a given fluctuation. MFT applies to systems described at the mesoscopic level by a (fluctuating) continuity equation of the form
\be
\partial_t\rho + \partial_x j = 0 \, ,
\label{hydro}
\ee
where $\rho(x,t)$ and $j(x,t)$ are the density and current fields, respectively, and $t$ and $x\in[0,1]$ are the macroscopic time and space variables, obtained after a diffusive scaling limit such that $x=i/N$ and $t=\tilde{t}/N^2$, with $i$ and $\tilde{t}$ the microscopic space and time variables. Periodic boundary conditions, the case of interest here, thus imply $\rho(0,t)=\rho(1,t)$ and $j(0,t)=j(1,t)$. Moreover, as the system is isolated, the total density remains constant
\be
\rho_0 = \int_0^1 \rho(x,t) dx \, .
\label{Arho0}
\ee
The current field in eq. (\ref{hydro}) is in general a fluctuating quantity, and can be written as
\begin{equation}
j(x,t)=-D(\rho)\partial_x \rho(x,t) +\sigma(\rho) E + \xi(x,t).
\label{current}
\end{equation}
The first term is Fick's (or equivalently Fourier's) law, where $D(\rho)$ is the diffusivity (which might be a nonlinear function of the local density). The second term is just the coupling to the external field $E$, mediated by the so-called mobility $\sigma(\rho)$, and $\xi(x,t)$ is the current noise that is gaussian and white,
\begin{equation}
\langle \xi(x,t)\rangle=0, \qquad \langle \xi(x,t) \xi(x',t')\rangle = \frac{\sigma(\rho)}{N} \delta(x-x') \delta(t-t'),
\label{noise}
\end{equation}
This gaussian fluctuating field is expected to emerge for most situations in the appropriate mesoscopic limit as a result of a central limit theorem: although microscopic interactions for a given model can be highly complicated, the ensuing fluctuations of the slow hydrodynamic fields result from the sum of an enormous amount of random events at the microscale which give rise to gaussian statistics, with an amplitude of the order of $N^{-1/2}$, in the mesoscopic regime in which eq. (\ref{hydro}) emerges. For long times, a system described by the above set of equations reaches a nonequilibrium steady state characterized by a homogeneous density distribution $\rho_0$ and a nonzero net average current $\la q\ra = \sigma(\rho_0) E$. Note that, for the WASEP, the two essential transport coefficients $D(\rho)$ and $\sigma(\rho)$, which determine the complete macroscopic fluctuating behavior of the system, are $D(\rho)=\frac{1}{2}$ and $\sigma(\rho)=\rho(1-\rho)$ \cite{wasep}. In what follows we describe the theory in general, only particularizing for the WASEP case in the last stages of the calculation.

A simple path integral calculation starting from eq. (\ref{hydro}) then shows that the probability of a given history or path $\{\rho,j\}_0^{\tau}$ in mesoscopic phase space (i.e. the space spanned by the hydrodynamic fields) obeys a large deviation principle of the form $P(\{\rho,j\}_0^{\tau})\sim \exp(-N{\cal I}_{\tau}^E[\rho,j])$, where the rate function is given by \cite{Bertini,Pablo1,IFR}
\be
{\cal I}_{\tau}^E[\rho,j] = \int_0^{\tau}dt\int_0^1dx \frac{[j+D(\rho)\partial_x\rho-E \sigma(\rho)]^2}{2\sigma(\rho)} \, .
\label{Ipath}
\ee
We are interested here in the fluctuations of the space and time integrated current
\be
q=\frac{1}{\tau} \int_0^\tau dt\int_0^1 dx j(x,t) \, .
\label{Acurrent}
\ee
The probability of observing a given $q$ can now be written as a path integral over all possible histories $\{\rho,j\}_0^{\tau}$, weighted by its probability measure $P(\{ j,\rho\}_0^{\tau})$, and restricted to those histories compatible with the value of $q$ and $\rho_0$ in eqs. (\ref{Acurrent}) and (\ref{Arho0}), respectively, and the continuity equation (\ref{hydro}) at every point of space and time. For long times and large system sizes, this path integral is dominated by the associated saddle point and scales as $P(q)\sim \exp\{+\tau N G(q)\}$, where $G(q)$ is the current large deviation function (LDF) given by
\beq
G(q)=-\lim_{\tau \rightarrow \infty}\left[\frac{1}{\tau}\min_{\{\rho,j\}_0^{\tau}}{\cal I}_{\tau}^E (\rho,j)\right]
\label{LDF1}
\eeq
The optimal density and current fields solution of this variational problem, $\rho_q(x,t)$ and $j_q(x,t)$, can be interpreted as the optimal path the system follows in order to sustain a long-time current fluctuation. Finding the optimal fields is in general a complex spatiotemporal problem whose solution remains challenging in most cases. The problem becomes much simpler however in different limiting cases. For instance, one expects that small current fluctuations around the average, $q\simeq \la q \ra$, result from the random superposition of weakly-correlated local fluctuations of the microscopic jump process. In this case it is reasonable to assume the optimal density field to be just the flat, steady-state one,  $\rho_q(x,t)=\rho_0$, and hence $j_q(x,t)=q$, resulting in a simple quadratic form for the current LDF
\be
G_{\text{flat}}(q) = - \frac{(q-\sigma(\rho_0) E)^2}{2\sigma(\rho_0)} \, .
\label{gflat0}
\ee
Therefore Gaussian statistics is obtained for small (i.e. typical) current fluctuations, in agreement with the central limit theorem. The previous argument, revolving around small fluctuations, breaks down however for moderate current deviations where correlations may play a relevant role. In fact, Bodineau and Derrida have shown recently \cite{BD} that the flat profile indeed becomes unstable, in the sense that $G(q)$ increases by adding a small time-dependent periodic perturbation to the otherwise constant profile, whenever 
\be
8 \pi^2D^2(\rho_0)\sigma(\rho_0)+(E^2\sigma^2(\rho_0)-q^2)\sigma''(\rho_0)<0 \, ,
\label{Acond}
\ee
where $\sigma''$ denotes second derivative. This condition implies a well-defined critical current
\be
|q_c|=\sqrt{\frac{8\pi^2D^2(\rho_0)\sigma(\rho_0)}{\sigma''(\rho_0)}+E^2\sigma^2(\rho_0)} \, .
\label{Aqc}
\ee
In the WASEP case this implies that for any $|q|<|q_c|$ the instability emerges. This instability can be interpreted as a dynamical phase transition at the fluctuation level, and involves the spontaneous breaking of translation symmetry (see Fig. \ref{evolution}). In fact, the formation of a traveling wave corresponds to the emergence of a macroscopic jammed state which hinders transport of particles to facilitate a current fluctuation well below the average. Notice that, for the instability to exist, the strength of the driving field, $|E|$, must be large enough to guarantee a positive discriminant in eq. (\ref{Aqc}), namely
\be
|E|\geq |E_c| \equiv \text{Re}\left[\sqrt{-\frac{8\pi^2D(\rho_0)^2}{\sigma(\rho_0)\sigma''(\rho_0)}}\right] \, .
\label{AEc}
\ee
Therefore, since the mobility $\sigma(\rho)$ is positive definite, a non-zero threshold field only exists for models such that $\sigma''(\rho)<0$, which is the case of the WASEP here studied, where $\sigma(\rho)=\rho(1-\rho)$. Other transport models, as for instance the Kipnis-Marchioro-Presutti (KMP) model of heat conduction \cite{kmp,PabloSSB}, have $\sigma''(\rho)>0$ and hence $|E_c|=0$, thus exhibiting the aforementioned instability even in the absence of external fields \cite{PabloSSB}.

When the instability kicks in, an analysis of the resulting perturbation \cite{BD} suggests that the dominant form of the optimal profile is a traveling wave moving at constant velocity $v$
\be
\rho_q(x,t)=\omega_q(x-vt) \, ,
\label{Aonda}
\ee
which implies via the continuity equation (\ref{hydro})
\be
j_q(x,t)=q-v\rho_0+v\omega_q(x-vt) \, .
\label{q1}
\ee
Provided that the traveling-wave form remains as the optimal solution for currents well-below the critical threshold, the current LDF can now be written as
\begin{eqnarray}
G(q)=-\min_{\omega_q(x),v}\int_0^1\frac{dx}{2\sigma[\omega_q(x)]}[q-v\rho_0+v\omega_q(x)
\nonumber \\
+D[\omega_q(x)]\omega'_q(x)-\sigma[\omega_q(x)]E]^2 \, ,
\label{Aldfwave}
\end{eqnarray}
where we have dropped the time dependence due to the periodic boundary conditions, and the minimum is now taken over the traveling wave profile $\omega_q(x)$ and its velocity $v$. Expanding now the square in eq. (\ref{Aldfwave}), we notice that the terms linear in $\omega_q'$ give a null contribution due again to the system periodicity. Taking also into account the constraint $\int_0^1\omega_q(x) dx=\rho_0$, see eq. (\ref{Arho0}), one gets
\begin{eqnarray}
G(q)=-\min_{\omega_q(x),v}\left[\int_0^1dx (X(\omega_q)+\omega'_q(x)^2Y(\omega_q))\right]+qE,
\nonumber \\
\label{LDF3}
\end{eqnarray}
where, borrowing the notation of ref. \cite{BD},
\beq
X(\omega_q)=\frac{[q-v(\rho_0-\omega_q)]^2}{2\sigma(\omega_q)}+\frac{E^2\sigma(\omega_q)}{2}
\label{Xw}
\eeq
and
\beq
Y(\omega_q)=\frac{D(\omega_q)^2}{2\sigma(\omega_q)}.
\label{Yw}
\eeq
The differential equation for the optimal profile solution of the variational problem eq. (\ref{LDF3}) can be written as
\beq
X(\omega_q)-\omega'_q(x)^2Y(\omega_q)=C_1+C_2\omega_q \, .
\label{optprof}
\eeq
This equation generically yields a symmetric optimal profile with a $\omega_q(x)$ with a single minimum $\omega_1=\omega_q(x_1)$ and a single maximum $\omega_0=\omega_q(x_0)$ such that $|x_0-x_1|=1/2$ \cite{why}. The constants $C_1$ and $C_2$ can be expressed in terms of 
the extrema $\omega_1$ and $\omega_0$.

The optimal velocity also follows from the above variational problem,
\beq
v=-q\frac{\displaystyle\int_0^1dx\frac{(\omega_q-\rho_0)}{\sigma(\omega_q)}}{\displaystyle\int_0^1dx\frac{(\omega_q-\rho_0)^2}{\sigma(\omega_q)}}.
\label{eq1}
\eeq
It is worth emphasizing that the optimal velocity is proportional to $q$. This implies that the optimal profile solution of eq. (\ref{optprof}) depends exclusively on $q^2$ and not on the current sign, reflecting the Gallavotti-Cohen time-reversal symmetry. This invariance of the optimal profile under the transformation $q\leftrightarrow -q$ can now be used in eq. (\ref{LDF3}) to show explicitly the GC symmetry $G(q)-G(-q)=2Eq$. This fluctuation relation is fully confirmed in the simulations discussed in the main text. 

The constants $C_1$ and $C_2$ appearing in eq. (\ref{optprof}) can be expressed in terms of the extrema $\omega_1$ and $\omega_0$ of the profile via
\begin{eqnarray}
X(\omega_1)=C_1+C_2\omega_1 \, ,
\label{eq2} \\
X(\omega_0)=C_1+C_2\omega_0 \, .
\label{eq3}
\end{eqnarray}
Moreover, the extrema locations are fixed by the constraints on the distance between them and the total density of the system,
\begin{equation}
\frac{1}{2}=\int_{x_1}^{x_0}dx
=\int_{\omega_1}^{\omega_0}\sqrt{\frac{Y(\omega_q)}{X(\omega_q)-C_1-C_2\omega_q}}d\omega_q
\nonumber \\
\label{eq4}
\end{equation}
and
\begin{equation}
\frac{\rho_0}{2}=\int_{x_1}^{x_0}\omega_q(x) dx=\int_{\omega_1}^{\omega_0}\sqrt{\frac{\omega_q^2 Y(\omega_q)}{X(\omega_q)-C_1-C_2\omega_q}}d\omega_q \, ,
\label{eq5}
\end{equation}
where we have used in the last equality of both expressions the differential equation (\ref{optprof}). In this way, for fixed values of the current $q$ and the density $\rho_0$ (provided externally), we use eqs. (\ref{eq1})-(\ref{eq5}) in order to determine the five constants $\omega_1,\omega_0,C_1,C_2,v$ which can be used in turn to obtain the shape of the optimal density profile $\omega_q(x)$ from eq. (\ref{optprof}).

Notice that the unknown variables $\omega_0$, $\omega_1$ appear as the integration limits in eqs. (\ref{eq4}) and (\ref{eq5}), making this problem remarkably difficult 
to solve numerically. In what follows we show how, by performing a suitable change of variables, the integrals involved in the calculation can be transformed into known functions, as e.g. elliptic integrals of the first kind, thus allowing to derive an explicit analytical expression for $\omega_q(x)$ as a function of the relevant constants. We start by doing a change of variables to express all the relevant magnitudes in dimensionless form
\beq
v\equiv \frac{q}{\rho_0}u\, ;~~~~\omega_q(x)\equiv \rho_0 h(x) \, ;~~~~E\equiv \frac{q}{\rho_0^2}\epsilon \, .
\label{apc:cdv}
\eeq
Particularizing now our calculation for the WASEP, where the transport coefficients are $D(\rho)=1/2$ and $\sigma(\rho)=\rho(1-\rho)$, the differential equation (\ref{optprof}) for the traveling wave reads
\beq
\begin{array}{ccc}
\displaystyle h'(x)=\frac{2q}{\rho_0}\left[ (1-u+uh)^2-2D_2 h^2 (1-\rho_0 h)-\right. & \\
\displaystyle\left.2D_1 h (1-\rho_0 h)+\frac{\epsilon^2}{\rho_0^2}h^2(1-\rho_0 h)^2 \right]^{1/2},
\label{apc:hprima}
\end{array}
\eeq
where we have defined $D_i\equiv \frac{C_i \rho_0^i}{q^2}$, $i=1,\, 2$. Moreover, eqs. (\ref{eq2})-(\ref{eq3}) can now be written as
\begin{eqnarray}
(1-u+uh_k)^2 & = & 2D_2h_k^2(1-\rho_0 h_k)+2D_1h_k(1-\rho_0 h_k) \nonumber \\
 & - & \displaystyle\frac{\epsilon^2}{\rho_0^2}h_k^2 (1-\rho_0 h_k)^2 \, ,
\label{apc:eq2_3}
\end{eqnarray}
where $h_k=\omega_k/\rho_0$, with $k=0,\, 1$, are the extrema of the dimensionless profile $h(x)$. We can now use the above equations to write the constants $D_1$ and $D_2$ as a function of the dimensionless variables $h_1,~h_0$, $\epsilon$ and $u$, 
\begin{eqnarray}
\begin{array}{ccc}
D_2=\displaystyle\frac{1-u}{2(1-\rho_0 h_0)(1-\rho_0 h_1)}\left[ \frac{(1-u)}{h_1h_0}[1-\rho_0 (h_0+h_1)]\right. & \\
& \\
\displaystyle+\left.\frac{u^2}{(1-u)}+2u\rho_0 \right] +\displaystyle\frac{\epsilon^2}{2\rho_0^2}[1-\rho_0 (h_0+h_1)],
\label{apc:D2}
\end{array}
\end{eqnarray}

\beq
D_1=\frac{(1-u+uh_0)^2}{2h_0(1-\rho_0 h_0)}-D_2 h_0+\frac{\epsilon^2}{\rho_0^2}\frac{h_0 (1-\rho_0 h_0)}{2}.
\label{apc:D1}
\eeq
The remaining task consists in obtaining the three unknown variables $h_1$, $h_0$ and $u$ from eqs. (\ref{eq1}), (\ref{eq4}) and (\ref{eq5}). In particular, eq. (\ref{eq4}) boils down to
\begin{eqnarray}
\begin{array}{ccc}
\displaystyle\int_{h_1}^{h_0}dh[ (1-u+uh)^2-2D_2 h^2 (1-\rho_0 h)- & \\
\displaystyle2D_1 h (1-\rho_0 h)+\frac{\epsilon^2}{\rho_0^2}h^2(1-\rho_0 h)^2 ]^{-1/2}=\frac{q}{\rho_0}. 
\label{apc:eq4_4} & \\
\end{array}
\end{eqnarray}
The integrand of the above expression can be written in the following product form
\beq
(-ah+b)(h+c)(h-h_1)(h_0-h),
\label{apc:prodf}
\eeq
where the coefficients $a$, $b$ and $c$ (obtained by matching order by order) are (simple) functions of the unknown $h_1$, $h_0$, $u$ and the known $q$, $\rho_0$ and $\epsilon$. In order to eliminate the unknown extrema $h_1$, $h_0$ from the integration limits in eq. (\ref{apc:eq4_4}), we perform the following change of variables
\beq
h=h_0-\alpha(h_0-h_1).
\label{apc:cdv2}
\eeq
which allows us to rewrite eq. (\ref{apc:eq4_4}) as
\begin{eqnarray}
\frac{q}{\rho_0} & = & \int_0^1 \frac{d\alpha}{(h_0-h_1)\sqrt{a}}\Big[ \((\frac{h_0+c}{h_0-h_1}-\alpha\))  \nonumber \\
& \times &(1-\alpha)\alpha \((\alpha-\frac{ah_0-b}{a(h_0-h_1)}\)) \Big]^{-1/2}
\label{apc:eq4_5}
\end{eqnarray}
Defining now
\beq
\eta^2\equiv \frac{(ac+b)(h_0-h_1)}{(-ah_1+b)(h_0+c)}~~~\text{and}~~~z\equiv \frac{h_0-h_1}{\eta^2h_1}
\label{apc:defs}
\eeq
we get that Eq. (\ref{apc:eq4_5}) turns into
\begin{eqnarray}
\displaystyle\frac{q}{\rho_0} & = & \frac{2}{\sqrt{(ac+b)zh_1}}\int_0^1 \frac{d\alpha}{\sqrt{(1-\alpha^2)(1-\eta^2\alpha^2)}} \nonumber \\
 & = & \displaystyle\frac{2}{\sqrt{(ac+b)zh_1}} K(\eta^2)
\label{apc:eq4_6}
\end{eqnarray}
where $K(\eta^2)$ is the complete elliptic integral of the first kind.  It is worth emphasizing that $a,~b,~c,~z,$ and $\eta^2$ depend on $h_1,~h_0,~u$ and on $q,~\rho_0$ and $\epsilon$.

In a similar way, we can derive an expression for the adimensional optimal profile $h(x)$ by writing
\beq
\int_0^x d{\tilde x}=x=\int_{h_1}^{h}\frac{d{\tilde h}}{{\tilde h}'} \, ,
\label{apc:perfint1}
\eeq
and proceeding in the same way as before. This yields
\beq
\frac{qx}{\rho_0}=\frac{1}{\sqrt{(ac+b)zh_1}}\int_{\gamma}^1\frac{d{\alpha}}{\sqrt{(1-{\alpha}^2)(1-\eta^2{\alpha}^2)}}
\label{apc:perfint2}
\eeq
which is equivalent to 
\beq
\frac{qx}{\rho_0}=\frac{K(\eta^2)-F[\sin^{-1}(\gamma),\eta^2]}{\sqrt{(ac+b)zh_1}} \, ,
\label{apc:perfint3}
\eeq
where 
\be
\gamma(x) \equiv \displaystyle\sqrt{\frac{(h_0+c)[h(x)-h_1]}{(h_0-h_1)[h(x)+c]}}
\label{gamma}
\ee
and 
\be
F[\sin^{-1}(\gamma),\eta^2]\equiv \displaystyle\int_{0}^{\gamma}\frac{d{\alpha}}{\sqrt{(1-{\alpha}^2)(1-\eta^2{\alpha}^2)}} \nonumber
\ee
is the incomplete elliptic integral of the first kind. Now, by using eqs. (\ref{apc:eq4_6}) and (\ref{apc:perfint3}) we deduce that
\beq
F[\sin^{-1}(\gamma),\eta^2]=K(\eta^2)(1-2x) \, .
\label{apc:perfint4}
\eeq
Solving for $\gamma$ we obtain
\beq
\gamma(x)=\text{JacobiSN}\[[ K(\eta^2)(1-2x) \]] \, ,
\label{apc:gamma}
\eeq
where JacobiSN is the inverse of the incomplete elliptic integral of the first kind. This yields finally the optimal density profile, $\omega(x)=\rho_0 h(x)$, with $h(x)$ obtained from the above equation after taking into account eq. (\ref{gamma})
\beq
h(x)=\frac{h_1+c \Upsilon(x)}{1-\Upsilon(x)} \, ,
\label{apc:perfint5}
\eeq
with $\Upsilon (x) \equiv \displaystyle \left( \text{JacobiSN}\left[ K(\eta^2)(1-2x) \right] \right)^2\frac{h_0-h_1}{h_0+c}$.

Equation (\ref{apc:perfint5}) for the optimal traveling wave reflects the explicit dependence of the wave profile on the constants $h_1,~h_0$ and $u$. The remaining job consists in solving numerically for these constants in a self-consistent manner, once the explicit dependence of the extrema has been removed from integral limits. These constants can be thus obtained from eqs. (\ref{eq1}), (\ref{eq4}) and (\ref{eq5}), which can be written as
\beq
\int_0^{\frac{1}{2}} dx \frac{(h(x)-1)}{h(x)^2}(1-u+uh(x))=0,
\label{apc:eq1_7}
\eeq
\beq
\frac{2}{\sqrt{(ac+b)h_1z}}K(\eta^2)=\frac{q}{\rho_0},
\label{apc:eq4_7}
\eeq
\beq
\int_0^{\frac{1}{2}}h(x)dx=\frac{1}{2},
\label{apc:eq5_7}
\eeq
where $a$, $b$, $c$, $\eta^2$ and $z$ are known functions of $h_1,~h_0,~u$. In this way, for given values of $q$, $\rho_0$ and $E$, we get self-consistently $h_1,~h_0,~u$ using the form of the profile obtained in eq. (\ref{apc:perfint5}), which depends explicitly on these constants. 

To obtain the current LDF, $G(q)$, we just integrate numerically its expression (\ref{LDF3}) once particularized for the WASEP ($\sigma(\omega)=\omega (1-\omega),~D(\omega)=\frac{1}{2}$), using the optimal wave profile and velocity obtained from the previous calculation. Finally, to compute the Legendre transform of the current LDF, we just evaluate numerically $\mu(\lambda)=\max_q[\lambda q+G(q)]=\lambda q^*+G(q^*)$ with $q^*(\lambda)$ solution of the following equation
\beq
\lambda=-\displaystyle\left. \frac{\partial G(q)}{\partial q}\right|_{q=q^*}=\int_0^1 dx \frac{q^*(\lambda)-v(\rho_0-\omega(x))}{\omega(x)(1-\omega(x))}-E \, . \nonumber
\eeq


\begin{thebibliography}{99}
\bibitem{note1} Note that these rates converge for large $N$ to the standard ones found in literature, namely $\frac{1}{2}(1\pm\frac{E}{N})$, but avoid problems with negative rates for small $N$. In any case, the hydrodynamic description of both variants of the model is identical.
\bibitem{why} For the WASEP, where $D(\rho)=1/2$ and $\sigma(\rho)=\rho(1-\rho)$, the optimal traveling wave profile solution of eq. (\ref{optprof}) must be symmetric because this differential equation remains invariant under the transformation $x\to 1-x$. Periodicity implies that profile extrema, if any, come in pairs (maximum and minimum). Moreover, the profile has at most a single pair of extrema because, when we make $w'_q(x)=0$ in eq. (\ref{optprof}), the resulting equation is third order for WASEP. This, together with the symmetry of the profile, implies in turn that the extrema $\omega_1=\omega_q(x_1)$ and $\omega_0=\omega_q(x_0)$ are such that $|x_0-x_1|=1/2$.
\end{thebibliography}
\end{document}